\documentclass[twocolumn,floats,prb,aps,showpacs]{revtex4}
\usepackage[final]{graphicx}
\usepackage[dvipsnames]{color}
\usepackage{dcolumn}
\usepackage{hhline}
\usepackage{braket}
\usepackage{amsmath}
\DeclareGraphicsExtensions{.pdf,.eps,.jpg,.png}

\begin{document}

\title{Exploring approximations to the $GW$ self-energy ionic gradients}

\author{C. Faber$^{1,2,3}$, P. Boulanger$^{1,2}$, C. Attaccalite$^{1,2}$, E. Cannuccia$^{4,5}$, I. Duchemin$^{2,3}$, T. Deutsch$^{2,3}$, X. Blase$^{1,2}$}

\affiliation{ 
$^{1}$CNRS, Institut N\'{e}el, 38042 Grenoble, France, \\
$^{2}$Universit\'{e} Grenoble Alpes, 38042 Grenoble, France, \\
$^{3}$INAC, SP2M/L$\_$sim, CEA cedex 09, 38054 Grenoble, France.\\
$^{4}$Institut Laue Langevin, BP 156, 38042 Grenoble, France, \\
$^{5}$Laboratoire de Physique des Interactions Ioniques et Mol\'eculaires, Aix-Marseille Universit\'e/CNRS, Campus de Saint J\'er\^ome,  13397 Marseille, France.
}

\date{\today}

\begin{abstract}
The accuracy of the many-body perturbation theory $GW$ formalism to calculate electron-phonon coupling
matrix elements has been recently demonstrated in the case of a few important systems. However, the
related computational costs are high and thus represent strong limitations to its widespread application. In the present
study, we explore two less demanding alternatives for the calculation of electron-phonon coupling matrix elements on the
many-body perturbation theory level. Namely, we test the accuracy of the static Coulomb-hole plus screened-exchange (COHSEX)
approximation and further of the constant screening approach, where variations of the screened 
Coulomb potential $W$ upon small changes of the atomic positions along the vibrational eigenmodes are neglected. 
We find this latter approximation to be the most reliable, whereas the static COHSEX ansatz leads to substantial errors.  Our 
conclusions are validated in a few paradigmatic cases: diamond, graphene and the $C_{60}$ fullerene. These findings 
open the way for combining the present many-body perturbation approach with efficient linear-response theories. 
\end{abstract}


\maketitle


\section{Introduction}

Electron-phonon coupling (EPC) occupies a prominent role in various fields of condensed matter 
physics, including phonon-mediated superconductivity, photoemission band gap renormalization, current 
carriers inelastic scattering or the lifetime of hot electrons. Concerning the calculation of EPC matrix 
elements on the \textit{ab initio} level, up to now, mainly density functional theory (DFT) \cite{dft}
and its perturbative linear response extensions (DFPT) \cite{dfpt1,dfpt2} have been applied,
providing important information at the microscopic level. 


Recently, several studies questioned the accuracy of the DFT-based EPC matrix elements  when obtained with (semi)local functionals such as LDA or PBE. \cite{PBE}
By way of example, the electron-phonon coupling involving specific electronic and phonon modes in graphene\cite{Lazzeri08,Basko08} and graphite, 
\cite{gruneis2009phonon} the value of the electron-phonon coupling potential to states at the Femi level in the electron-doped 
fullerene, \cite{Janssen10,Faber11} superconducting bismuthates and transition-metal chloronitrides,
 \cite{Yin03} or the renormalization of the photoemission band structure of pentacene \cite{Ciuchi12} 
and diamond crystals, \cite{Antonius14} were shown to be affected by a significant underestimation of 
the EPC matrix elements when calculated within DFT and (semi)local functionals. 

As a cure to such problems, many-body perturbation theory (MBPT) techniques within the so-called $GW$
approximation \cite{Schwinger64,Hedin65,Hybertsen86,Godby88,Onida02,Aryasetiawan98}
showed a clear improvement when compared to available experimental data. \cite{Lazzeri08,gruneis2009phonon,Faber11,Yin03,Ciuchi12,Antonius14} 
Unfortunately, the kind of theories that are available within DFT, and in particular the powerful DFPT formalism, \cite{dfpt1,dfpt2} are not yet
available within the framework of MBPT. 
Existing $GW$ calculations of EPC matrix elements have been therefore based on the 
frozen-phonon approach, necessitating a stepwise displacement of the atoms along the phonon modes with an explicit evaluation of the electronic
structure on the $GW$ level for each step. This demanding approach cannot be reconciled with very large unit cells and is thus only feasible for
zone-center or zone-boundary phonon modes.
Going beyond the frozen-phonon approach, in order to access EPC matrix elements on the $GW$ level for a large
number of electron and phonon wave vectors, remains a considerable
challenge to the \textit{ab initio} community. Significant work is still needed to come up with a scheme
that may allow to generalize the pioneering above-mentioned studies to a much larger set of systems.

In the present study, we explore the merits of the static coulomb-hole plus screened exchange (COHSEX) 
\cite{Hedin65,Hybertsen86,Farid,BrunevalThesis,Bruneval06a,Gatti07,Bruneval06b,Trani10,Vidal10,Lucero11,Kang10}
approximation to the $GW$ self-energy for the calculation of the electron-phonon coupling matrix 
elements in the isolated fullerene molecule, in diamond and in graphene. Our definitions of the relevant 
EPC  matrix  elements follow previous studies related to the superconducting transition in the fullerides, \cite{Janssen10,Faber11} 
the zero-point motion renormalization of the gap in diamond \cite{Antonius14} and the gap opening
through electron-phonon coupling in graphene, \cite{Lazzeri08} respectively. For the same set of systems, we also explore the accuracy 
of the constant-screening approximation, namely the assumption that the screened Coulomb 
potential can be considered, to linear order, as a constant upon small changes of the atomic positions. Whereas COHSEX leads to non-negligible
discrepancies, this latter approximation will be shown to be robust and accurate.
The present results offer important perspectives in making the $GW$ 
formalism amenable to the study of condensed matter phenomena involving electron-phonon coupling.


\section{Methodology}

We briefly introduce the many-body perturbation theory Green's function formalism, providing a solid 
framework for the calculation of quasiparticle energies $E$. In such an approach, the one-body quasiparticle 
eigenvalue equation reads:

\begin{eqnarray*}
   \left( { -{\nabla}^2 \over 2} + V^{ion}(\textbf{r}) \right) \phi(\textbf{r}) + 
       V^{H}(\textbf{r}) \phi(\textbf{r})   \\
   + \int d\textbf{r}' \; \Sigma(\textbf{r},\textbf{r}'; E ) \phi(\textbf{r}') = 
       E \phi(\textbf{r}), 
\end{eqnarray*}

\noindent where $V^{ion}$ and $ V^{H}$ are the ionic and Hartree potential, 
respectively. The self-energy $\Sigma(\textbf{r},\textbf{r}'; E )$ replaces the well-known
exchange-correlation potential of density functional theory or the exchange operator in 
the Hartree-Fock formalism.  In general, it is non-local, energy-dependent and non-Hermitian. 
Derived within Schwinger's functional derivative approach to perturbation theory,
\cite{Schwinger64} the $GW$ approximation to the self-energy leads to:

\begin{eqnarray*}
 \Sigma^{GW}({\bf r},{\bf r}'; E ) = {i \over 2\pi} \int d\omega e^{i \delta \omega} W({\bf r},{\bf r}'; \omega )
             G({\bf r},{\bf r}'; E + \omega),
\end{eqnarray*}

\noindent where $G$ and $W$ are the time-ordered one-particle Green's function and the dynamically 
screened Coulomb potential, respectively, and $\delta=0^+$ a small positive infinitesimal.

The  Coulomb-hole plus screened-exchange (COHSEX) representation of the $GW$ self-energy was originally
introduced using a time representation of $G$, $W$ and $\Sigma$. \cite{Hedin65,Bruneval06a,BrunevalThesis}
We follow Hybertsen and Louie \cite{Hybertsen86} by using the following spectral representations of $G$ and $W$:

\begin{eqnarray*}
G({\bf r},{\bf r}'; E+ \omega) &=& \sum_n { \phi_n({\bf{r}}) \phi_n^*({\bf{r}}') \over
 E+ \omega - \varepsilon_n - i\delta \text{sgn}(\varepsilon_n-\mu) } \\
W({\bf r},{\bf r}'; \omega ) &=& v({\bf r},{\bf r}') +
      \int_0^{\infty} d\omega' { 2\omega' B({\bf r},{\bf r}'; \omega') \over {\omega}^2 - (\omega'-i \delta)^2 }, \nonumber
\end{eqnarray*}

\noindent where $G$ has been written in terms of one-body eigenstates $\phi_n$ and eigenenergies $\varepsilon_n$, typically starting DFT Kohn-Sham 
or Hartree-Fock  solutions, and $W$ in terms of its spectral function $B$. These expressions 
allow to  obtain the pole structure of both $G$ and $W$. From the residue theorem, one then rapidly obtains 
an \textit{exact} decomposition of $\Sigma$:

\begin{eqnarray*}
\label{eq:Dyn_SEX}
\Sigma^{SEX}({\bf r},{\bf r}';E)  &=& - \sum_n^{occ} W({\bf{r}},{\bf{r}}'; E - \varepsilon_n) \phi_n({\bf{r}}) \phi_n^*({\bf{r}}') \\ \nonumber
\Sigma^{COH}({\bf r},{\bf r}';E)  &=&  \sum_n  \phi_n({\bf{r}}) \phi_n^*({\bf{r}}') \mathcal{P}
         \int_0^{\infty} d\omega' { B(r,r';\omega') \over E - \varepsilon_n - \omega' },
\end{eqnarray*}

\noindent  where $\mathcal{P}$ indicates the principal value. ${\Sigma}^{SEX}$, which involves a summation 
over the occupied states only, originates from the poles of $G$. It is called the screened exchange interaction in 
analogy to the bare exchange term that can be obtained by replacing $W$ with the energy-independent 
bare Coulomb potential $v$.  ${\Sigma}^{COH}$ originates from the poles of $W$ and represents the  Coulomb-hole
contribution, since it can be shown to be related to the interaction of an electron with its related 
adiabatically built correlation hole.

The static approximation to the exact COHSEX decomposition assumes that 
$(E - \varepsilon_n) \simeq 0$ for all (n), leading to simplified static screened exchange and Coulomb-hole 
expressions, namely:

\begin{eqnarray*}
\label{eq:static_SEX}
\Sigma^{SEX}_{static}({\bf r},{\bf r}';0)  &=& - \sum_n^{occ} W({\bf{r}},{\bf{r}}'; 0) \phi_n({\bf{r}}) \phi_n^*({\bf{r}'}) \\
\Sigma^{COH}_{static}({\bf r},{\bf r}';0)  &=&  {1 \over 2} \sum_n \phi_n({\bf{r}}) \phi_n^*({\bf{r}}') 
{\widetilde W}({\bf{r}},{\bf{r}'}; 0)  \\
                 &=&  {1 \over 2} \delta({\bf{r}}-{\bf{r}'})  {\widetilde W}({\bf{r}},{\bf{r}}'; 0).
\end{eqnarray*}

\noindent Here, $\; {\widetilde W} = (W - v)$ is the difference between the screened and bare Coulomb potential.
As a result, besides being a static approximation, this Coulomb-hole term is also local in space.
Such a static COHSEX approximation (labeled COHSEX here below) was shown to yield too large gaps in the 
case of semiconductors.  \cite{Hybertsen86} By way of example, in the present case of the C$_{60}$ molecule, the COHSEX gap is found to be 
5.3 eV, i.e. 0.4 eV larger than the $\sim$4.9 eV experimental gap.\cite{EXP} 
Nevertheless, this has to be compared to the starting 1.6 eV DFT-LDA Kohn-Sham gap which is dramatically too small. 

While it cannot be claimed that the static COHSEX approach is a good approximation to absolute quasiparticle 
energies, we emphasize that we are interested in quasiparticle energy differences upon small (infinitesimal) 
atomic lattice motions. The main assumption on which we rely to calculate the electron-phonon coupling
within the COHSEX approximation is that the \textit{variations} of the dynamical contribution to the self-energy
can be neglected. This can be \textit{a priori} rationalized by emphasizing that dynamical interactions are
driven by the plasmons dynamics, collective excitations less sensitive to small atomic
displacements than single-particle excitation energies and wave functions. It remains, however,
that besides the static approximation, the spatially  local character of the static COH term is 
at odds with the nonlocality of the full $GW$ self-energy.

In order to further justify the following results concerning the constant screening approach, namely the second approximation 
we explore in this study, it is instructive to consider the  $GW$ plus Bethe-Salpeter 
formalism. \cite{Sham66,Hanke79,Strinati82} The latter is a many-body perturbation theory approach concerned with describing the linear response of a 
system with respect to a time-dependent external perturbation and thus the Bethe-Salpeter equation (BSE) is the MBPT analogue to time-dependent DFT. At
the heart of the $GW$/BSE approach  lies the variation $(\partial GW /  \partial \lambda)$, where the ``perturbation" 
$\lambda$ is the one-body Green's  function $G$. The most common approximation that has been shown to be remarkably accurate
\cite{Rohlfing98,Benedict98,Albrecht98} is to replace the $GW$ self-energy by its static COHSEX approximation, and to consider further that
$(\partial W /  \partial \lambda)=0$. It is such a simplified scheme we aim to explore in the present study, differing in the fact that
 the perturbation $\lambda$ is now induced by a vibrational distortion of the system.

\section{Technical details}

The many-body $GW$ and COHSEX calculations for the isolated $C_{60}$ fullerene are performed using the {\sc{Fiesta}} code, an 
implementation of the $GW$ formalism within a Gaussian basis. \cite{Fiesta1,Fiesta2,Fiesta3} We start from DFT-LDA eigenstates 
calculated with the {\sc{Siesta}} package \cite{Siesta} and a triple-zeta plus polarization (TZP) basis \cite{tzpbasis} for the 
description of the valence orbitals combined with standard norm-conserving pseudopotentials. \cite{Pseudo} We use the resolution 
of the identity technique (RI-SVS) with an even-tempered auxiliary basis set formed of four Gaussians per angular channel 
(up to \textit{d} character) with exponents between 0.2 and 3.2 Bohr$^{-2}$. 
With such a basis, our $G_0W_0$@LDA ionization potential and HOMO-LUMO  gap are  found to be 7.33 eV and 4.39 eV (B3LYP geometry), respectively, starting
from a 1.66 eV DFT-LDA gap. This is in close agreement with the 7.45 eV and 4.40 eV plane wave  $G_0W_0$@LDA values  of  Ref.~\onlinecite{galli}
(LDA geometry).  The  $G_0W_0$@LDA  gap  remains smaller than the $\sim$4.9 eV  experimental value, an issue that will be addressed by the  partially 
self-consistent scheme used in the present study, as discussed below.

Following Ref.~\onlinecite{Janssen10}, we use the relaxed structure and phonon eigenmodes generated within a DFT-B3LYP approach using a 6-311G(d) basis.
This was shown to provide vibrational frequencies in excellent agreement with Raman measurements. \cite{Janssen10,noteph} 
Concerning the calculation of the EPC matrix elements, our present study follows the approach described in Refs.~\onlinecite{Janssen10,Faber11}, where the
electron-phonon coupling strength associated with the (electron-doped) lowest unoccupied 3-fold degenerate $t_{1u}$ molecular orbital (LUMO) 
has been explored within DFT and $GW$. The specific choice of the LUMO level is dictated by the physics of the phonon-mediated superconducting 
transition in the fullerides. \cite{Hebard91,Gunnarsson97} We adopt the standard definition of the effective electron-phonon 
coupling potential $V^{ep}$ that enters e.g. the McMillan formula for the superconducting transition temperature through the 
dimensionless parameter  ${\lambda}=N(E_F)V^{ep}$, where $N(E_F)$ is the density of states at the Fermi level. Namely, $V^{ep}$ reads in the so-called molecular limit: 
\cite{Antropov93,Cote98}

\begin{equation}
V^{ep} = \frac{1}{9 M } \sum_{\nu} \frac{1}{\omega^2_{\nu}} \sum^{3}_{m=1} 
   \left| \frac{\partial \epsilon_m}{\partial \mathbf{u}^{\nu}} \right|^2. 
 \label{eq:Vep}
\end{equation}

\begin{figure*}
 \includegraphics[width=0.6\linewidth]{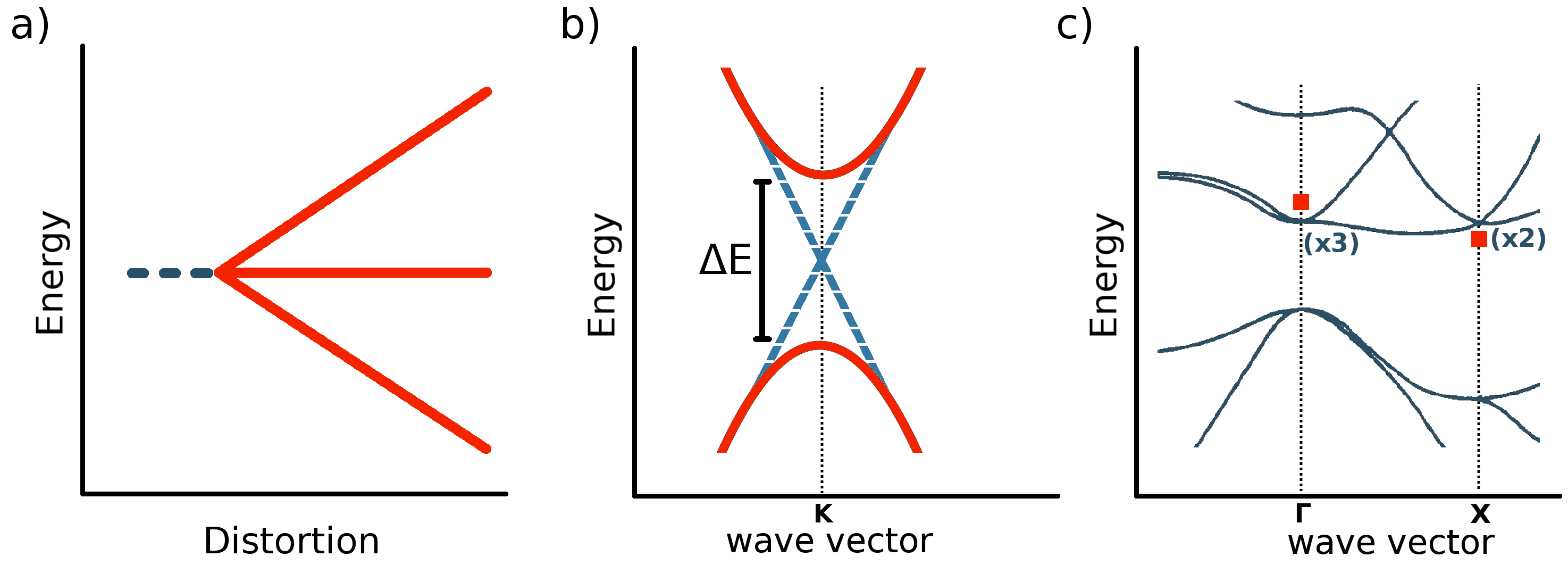}
 \caption{
Schematic representation of changes in the electronic structure under phonon distortion: a) in C$_{60}$, the 3-fold  LUMO levels
(blue) split under a distortion (red) along a $H_g$ phonon mode.  b) In graphene, an energy gap ${\Delta} E$ at the Dirac point  
is opened (red) when distorting the equilibrium structure (blue)  along the ${\Gamma}-A_1'$ or $K-E_{2g}$ phonon modes.  
c) In diamond, the  lowest lying conduction  bands at $\Gamma$/$X$ (blue) are renormalised by their coupling to a zone-boundary phonon mode.
For clarity, only the states  showing the largest shift are plotted (red dot, see text).  } 
\label{fig:methods}
\end{figure*}

\noindent
Here, $M$ is the mass of the carbon ions, $\omega_{\nu}$ is the 
frequency of the vibrational mode with index $(\nu)$  and $\mathbf{u}^{\nu}$ is the vibrational polarization vector. As already discussed before, the
electronic states with energy  $(\epsilon_m)$ are limited to the 3-fold 
degenerate LUMO  level. From group theory analysis it follows that only the $H_g$ and $A_g$
vibrational modes can couple to these states. A schematic representation of the $t_{1u}$ level splitting as a function of the deformation
amplitude along a $H_g$ mode is provided in Fig.~\ref{fig:methods}(a). The EPC matrix elements are consequently related to the slopes
(${\partial \epsilon_m}/{\partial \mathbf{u}^{\nu}}$), showing a strong dependence on the formalism adopted to calculate the energies $\epsilon_m$.
These energy derivatives are calculated within the frozen phonon approach, using a symmetric five points finite-difference formula.

Such an effective electron-phonon potential was calculated for $C_{60}$ in Refs.~\onlinecite{Janssen10,Faber11}, both within DFT using diverse (semi)local
and hybrid exchange-correlation functionals and $GW$.  While DFT-LDA values were shown to significantly underestimate the coupling energy, DFT calculations
performed with global hybrids containing 20$\%$-30$\%$ of exact exchange and the $GW$ approach were found \cite{Janssen10,Faber11} to compare favorably to
gas phase experimental data. \cite{Iwahara10,Hands08}  As a drawback of DFT calculations with hybrids, it was evidenced that the resulting $V^{ep}$ potential
would quickly increase with the amount of exact exchange, yielding the usual question of the proper choice of the functional parameters for a given system. 

In the present study, we are concerned with exploring simplified $GW$ schemes, i.e approximations reducing the needed computational effort. As such,
quasiparticle energies are evaluated at a single-shot COHSEX and a partially 
self-consistent evCOHSEX level, where the quasiparticle energies are reinjected self-consistently 
in the construction of $G$ and $W$, while keeping the starting Kohn-Sham LDA wave functions unchanged. Our reference point are partially self-consistent
ev$GW$ calculations with self-consistency on the eigenvalues as presented in Ref.~\onlinecite{Faber11} for the calculation of the EPC matrix elements in
$C_{60}$. Such a simplified approach to full self-consistency, justified by the dramatically too small starting point DFT-LDA gap for small isolated molecules,
has been shown in the case of gas phase organic systems to produce more accurate ionization  potentials and electronic affinities,
\cite{Fiesta1,Fiesta2,Hahn05,Tiago08,Sharifzadeh12,Hogan13} together with improved optical excitation energies at the  $GW$/BSE level, \cite{Fiesta3,Faber13,Baumeier13,Boulanger14}
when LDA or PBE Kohn-Sham eigenstates are chosen as starting points. Consistently, when starting from DFT-LDA, EPC matrix elements in $C_{60}$ were
also found to be in closer agreement with experiment within ev$GW$ as compared to $G_0W_0$. 
\cite{Faber11}  

\begin{table*}
\begin{tabular}{c|c|c|c|c|c}
\hline
 & LDA  &  $G_0W_0$ & $G_0W_0$($W$)  & COHSEX & COHSEX($W$)    \\

\hline
 \multicolumn{6}{c}{Graphene} \\
\hline
${\Gamma}$-$E_{2g}$   &   44.3   &       68.1 (65.2)    &  72.0   &  81.5      &    83.2             \\
${\bf K}$-A$_{1}^{'}$        &   88.6   &       201(187)   &  207   &    260    &      256        \\

\hline
 \multicolumn{6}{c}{Diamond} \\
\hline
${\bf\Gamma}_{15c} $  &    1.281     &   0.611   &   0.657   &  1.311    & 1.445   \\ 
${\bf X}_{1c}$        &   -2.009     &  -1.031   &  -1.066   & -2.124    & -2.273   \\
\hline
\end{tabular}
\caption{Calculated EPC-related quantities (see text) for graphene and diamond. In the case of graphene, these correspond to the splitting of the
degenerate highest valence and lowest conduction band at the Dirac point under the influence of a (${{\bf q}\nu}={\bf{\Gamma}}$-$E_{2g}$) and
(${{\bf q}\nu}={\bf K}$-A$_1^{'}$) phonon mode and are expressed in $(eV/\AA)^2$.  Number in parenthesis are obtained with a plasmon pole energy 
of 7 eV, close to the graphene $\pi$-plasmon energy, instead of the 27 eV originally used in Ref.~\onlinecite{Lazzeri08} (see text).  For  diamond, these are the 
results of equation (\ref{eq:EPC2}) for  the ${\bf X}_{1c}$ and ${\bf \Gamma}_{15c}$ conduction states and for the ${\bf X}_4$ zone-boundary phonon 
mode only (values in eV). This corresponds to  twice the zero-point renormalization arising from this given phonon mode (see Ref.~\onlinecite{Antonius14}).}
\label{table1}
\end{table*}

For the periodic diamond and graphene systems, we use the Yambo code which implements many-body perturbation theory within 
a plane wave formalism. \cite{Yambo} The starting DFT Kohn-Sham eigenstates are generated with the Abinit code.
\cite{Abinit} In the case of graphene, the running parameters are identical to those used in a previous $GW$ study 
of the electron-phonon coupling in this material, \cite{Lazzeri08} namely we use $a=2.46$~\AA\ as lattice parameter,  15 a.u. distance 
between the layers, \cite{notegraphene} $36 \times 36 \times 1$ {\bf k}-points for the $GW$ calculations in the Brillouin zone (BZ) 
corresponding to the primitive (2-atoms) cell and the nearest equivalent {\bf k}-grid  in the BZ corresponding to the supercell 
needed to describe zone-boundary (${\bf q}={\bf K}$) phonons  in the frozen-phonon approach. The energy cutoff on the plane wave 
basis was set to 60~Ry. All conduction bands within an energy range of 45~eV above the Fermi level were used to build $G$
and $W$, and a cutoff of 4 Ha for the construction of the dielectric function.  The Godby-Needs plasmon pole model was employed for the dynamical dielectric constant entering in $W$. 

As schematically described in Fig.~\ref{fig:methods}(b), we focus on the splitting $\Delta E$ of the degenerate occupied and unoccupied levels at the Dirac 
point caused by the coupling to the ${{\bf q}\nu}={\bf{\Gamma}}$-$E_{2g}$ optical mode at zone center together with the zone-boundary 
${{\bf q}\nu}={\bf K}$-A$_1^{'}$ phonon, for which the strongest renormalization of the coupling constant have been observed upon replacing 
the DFT-LDA approach by the $G_0W_0$ formalism. \cite{Lazzeri08,Basko08} 
The EPC matrix elements of interest are thus here related to the derivatives:  $(\partial \Delta E / \partial {\bf u}_{{\bf q}\nu})$,
namely (see Ref.~\onlinecite{Lazzeri08}):

\begin{equation}
  \langle D^2_{{\bf q}\nu} \rangle = {1 \over A_{{\bf q}\nu}} \left( \partial \Delta E \over \partial \mathbf{u}_{\nu\mathbf{q}}\right)^2,
\end{equation}

\noindent expressed in $(eV/\AA)^2$, where $A_{{\bf q}\nu}$ equals 16 and 8 for ${{\bf q}\nu}={\bf{\Gamma}}$-$E_{2g}$
and ${{\bf q}\nu}={\bf K}$-A$_1^{'}$, respectively.  We performed ``single-shot"$G_0W_0$ and COHSEX calculations on top of  DFT-LDA results, 
namely the standard perturbative (non self-consistent) $GW$ approach to obtain quasiparticle energies.

For diamond, following Antonius and coworkers, \cite{Antonius14}
we have used a relaxed DFT-LDA geometry (lattice parameter $a=3.591~\AA$) with a plane wave cutoff energy of 80~Ry and
norm-conserving pseudopotentials. All calculations were performed with a 8x4x4 uniform Monkhorst-Pack grid to describe the supercell necessary
for the frozen-phonon simulation of the (${\bf q} = X$) phonon. We have used 160 bands and a cutoff of 8 Ha for the construction of the dielectric function and
the $GW$ self-energy. Following Refs.~\onlinecite{Antonius14} and \onlinecite{AHC} where the diamond band-gap renormalization by the
zero-point motion was studied,  we define the relevant EPC matrix elements as the second derivative of the electronic eigenvalues with
respect to the phonon displacements at equilibrium:

\begin{equation}
 \frac{\partial \epsilon_{m\mathbf{k}}}{\partial n_{\nu\mathbf{q}}} = 
     \frac{\hbar}{2M\omega_{{\nu}\mathbf{q}}} \frac{\partial^2}{\partial \mathbf{u}^{2}_{\nu\mathbf{q}}} \epsilon_{m\mathbf{k}},
 \label{eq:EPC2}
\end{equation}

\noindent where $n_{\nu \mathbf{q}}$ is the occupation of the phonon state $(\nu)$ of wave vector $\mathbf{q}$. 
The EPC matrix elements are now related to the temperature dependent renormalization of the electronic bands.
We evaluate the second-derivative of the eigenvalues using a symmetric five-point finite-difference formula with displacements
of 0.01, 0.02 and 0.03 \AA. In Ref.~\onlinecite{Ponce14}, it was already shown that standard DFT and DFPT methods with (semi)local
functionals strongly underestimate 
(by ca. 30\%) the zero-point renormalization of the direct gap of diamond compared to experiment, while 
Ref.~\onlinecite{Antonius14} demonstrated that $G_0W_0$ calculations restored the good agreement between 
theory and experiment.  We focus here below on the renormalization of the lowest conduction band edge at 
($m{\bf k}={\bf X}_{1c}$) and ($m{\bf k}={\bf \Gamma}_{15c}$) by the coupling to the ($\nu{\bf q}={\bf X_4}$) phonon modes at
the zone-boundary (see Fig.~\ref{fig:methods}c). It is for such a phonon wave vector that the EPC matrix elements were found to be large and that also
the $GW$ correction was shown to be the most significant.\cite{Antonius14} 
Following Ref.~\onlinecite{Antonius14}, the provided EPC matrix elements are averaged over the degenerate
electronic manifold considered (namely the 3 degenerate levels at zone-center and the 2 degenerate levels 
at the ${\bf X}_{1c}$ zone boundary).

\begin{table*}
\begin{tabular}{c|c|c|c|c|c|c}
\hline
           &  \multicolumn{5}{c}{Theory}  &  {Experiment} \\
\hline
   Mode    & LDA  & ev$GW$ &   COHSEX & evCOHSEX & evCOHSEX($W$) &   \\

\hline
 \multicolumn{7}{c}{Fullerene} \\
\hline
  $H_g(1)$ & 4.7  &  5.85  &	4.99  & 6.3   &	6.3   & \\
  $H_g(2)$ & 9.3  &  10.1  &	9.99  &	10.3  &	11.9  & \\
  $H_g(3)$ & 8.8  &  12.95 &	11.96 &	13.7  &	15.3  & \\
  $H_g(4)$ & 4.2  &  5.5   &	4.9   &	5.4   &	5.7   & \\
  $H_g(5)$ & 3.98 &  4.9   &	4.5   &	5.3   &	5.5   & \\
  $H_g(6)$ & 1.8  &  1.99  &	2.05  &	2.2   &	2.4   & \\ 
  $H_g(7)$ & 15.8 &  25.4  &	23.3  &	27.9  &	26.5  & \\
  $H_g(8)$ & 13.1 &  18.3  &	17.1  &	19.8  &	19.5  & \\
  $A_g(1)$ & 1.2  &  1.9   &	1.85  &	1.8   &	1.2   & \\
  $A_g(2)$ & 7.2  &  13.98 &	13.2  &	15.4  &	12.2  & \\
\hline
Total $A_g$ & 8.4 &  15.9  &    15.0  &	17.2  & 13.4  & \\ 
Total $H_g$ & 61.7&  84.96 &	78.8  &	90.9  &	93.1  & 96.2$^b$,96.5$^c$  \\
Total       & 70.2&  100.9 (108.6$^a$) & 93.8 &	108.1 &	106.6 & 106.8$^b$  \\
  
\hline
\end{tabular}
\caption{Mode-resolved $V^{ep}_{\nu}$ coupling potential associated with the 3-fold degenerate C$_{60}$ LUMO and the corresponding $V^{ep}$ totals
(in meV). For group symmetry reasons, only the $H_g$ and $A_g$ vibrational modes couple to this electronic state.  \\
$^a$ Ref.~\onlinecite{Faber11}, \\
$^b$ Ref.~\onlinecite{Iwahara10}, Table V,  \\
$^c$ Ref.~\onlinecite{Hands08}. }
\label{table2}
\end{table*}

\section{Results}

\subsection{Reference $GW$ calculations}

We first reproduce for validation and for reference the previously published $GW$ calculations related to EPC matrix elements
in graphene, diamond and $C_{60}$.
Our results are compiled in Table~\ref{table1} for diamond and graphene and in Table~\ref{table2} for $C_{60}$, respectively. 
Our LDA and $G_0W_0$ values for graphene are close to those found in Ref.~\onlinecite{Lazzeri08}, namely e.g. 201 eV/\AA\ ($G_0W_0$ value, present study) 
for the largest matrix element with the (${\bf K}$-A$_{1}^{'}$) phonon, to be compared to 193 eV/\AA\   in the previous study.\cite{notegraphene}
The difference can be explained by an increased five points, instead of only two, finite-difference formula, and
by the increase of the dielectric matrix size energy cutoff,  from 2 to 4 Ha cutoff. We also verified that changing the 
 Godby-Needs plasmon model input finite frequency,  from the (default) 27 eV value in Ref.~\onlinecite{Lazzeri08} 
 to a 7 eV value  closer to the $\pi$-plasmon resonance in graphene, does not change significantly the calculated 
 EPC strength  (see numbers in parenthesis in the Table~\ref{table1}). 
Such  differences are negligible with respect to the more than 100$\%$ increase  as compared to the LDA value. 

For diamond, the EPC contribution associated with the ${m\bf k}={\bf \Gamma}_{15c}$ conduction states are in good agreement with 
the LDA and $G_0W_0$ results of Antonius and coworkers, \cite{Antonius14} i.e. we find a 11 meV difference with respect to their
EPC in the case of LDA, while we find a difference of 8 meV for the $G_0W_0$ results. \cite{NOTEEPC} These small variations
can be ascribed to the different pseudopotentials (Ref.~\onlinecite{Ponce14} showed that pseudopotentials can lead to errors up to 50 meV),
and to the different convergence parameters.

Since Ref.~\onlinecite{Antonius14} focused on the renormalization of the direct band gap at the zone-center, comparison with the 
present results for the phonon coupling to the ${\bf X}_{1c}$ electronic state cannot be made. Our findings are, however, very
consistent with what is observed for the ${\bf \Gamma}_{15c}$ state, namely a dramatic decrease of the coupling strength with 
the $X_4$ phonon mode which, again, is the dominant coupling mode. The coupling energy is indeed found to be reduced by 52$\%$ 
and 49$\%$ from LDA to $GW$ for the ${\bf \Gamma}_{15c}$ and ${\bf X}_{1c}$ states, respectively. Since the ${\bf X}_{1c}$ state 
is closer to the true conduction band minimum, this is a strong indication that zero-point motion renormalization of the indirect 
gap will certainly also be strongly affected by the $GW$ correction. The full study of such an effect is beyond the scope of the
present paper.

For $C_{60}$, the total ev$GW$ coupling potential is within 8$\%$  of that found by Ref.~\onlinecite{Faber11} as a result of 
the larger TZP basis used here. \cite{tzpbasis} Our ev$GW$ $V^{ep}$ potential is found to increase by 44$\%$ with respect to 
the corresponding LDA value, to be compared with the 48$\%$ increase obtained in Ref.~\onlinecite{Faber11} with a smaller basis.
This indicates the good convergence of the $GW$ correction to the LDA value. We note that while DFT and $GW$ calculations are 
known to converge very differently with respect to basis size, we compare in the present study approaches which are much more
similar, namely ``standard and approximated" $GW$ calculations, indicating certainly an even better convergence of the 
differences we are interested in. 

\subsection{The COHSEX approximation}

\begin{figure}
 \includegraphics[width=\linewidth]{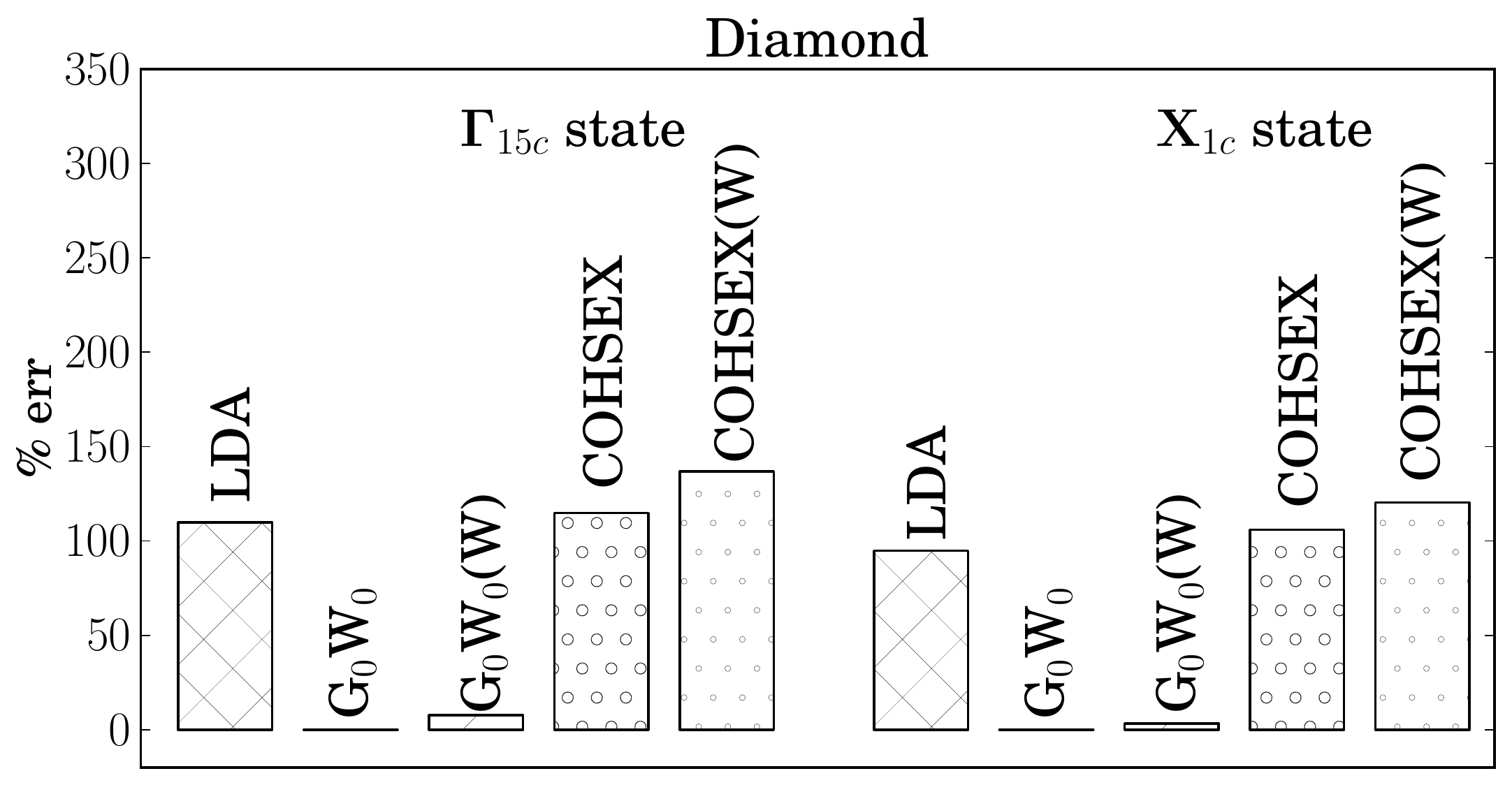}
 \caption{Relative error of the different approximations with respect to the $G_0W_0$ results for the ``thermal renormalization" EPC energies (Eqn.~\ref{eq:EPC2})
 for the lowest conduction band of diamond.}
  \label{figdiamond}
\end{figure}

\begin{figure}
 \includegraphics[width=\linewidth]{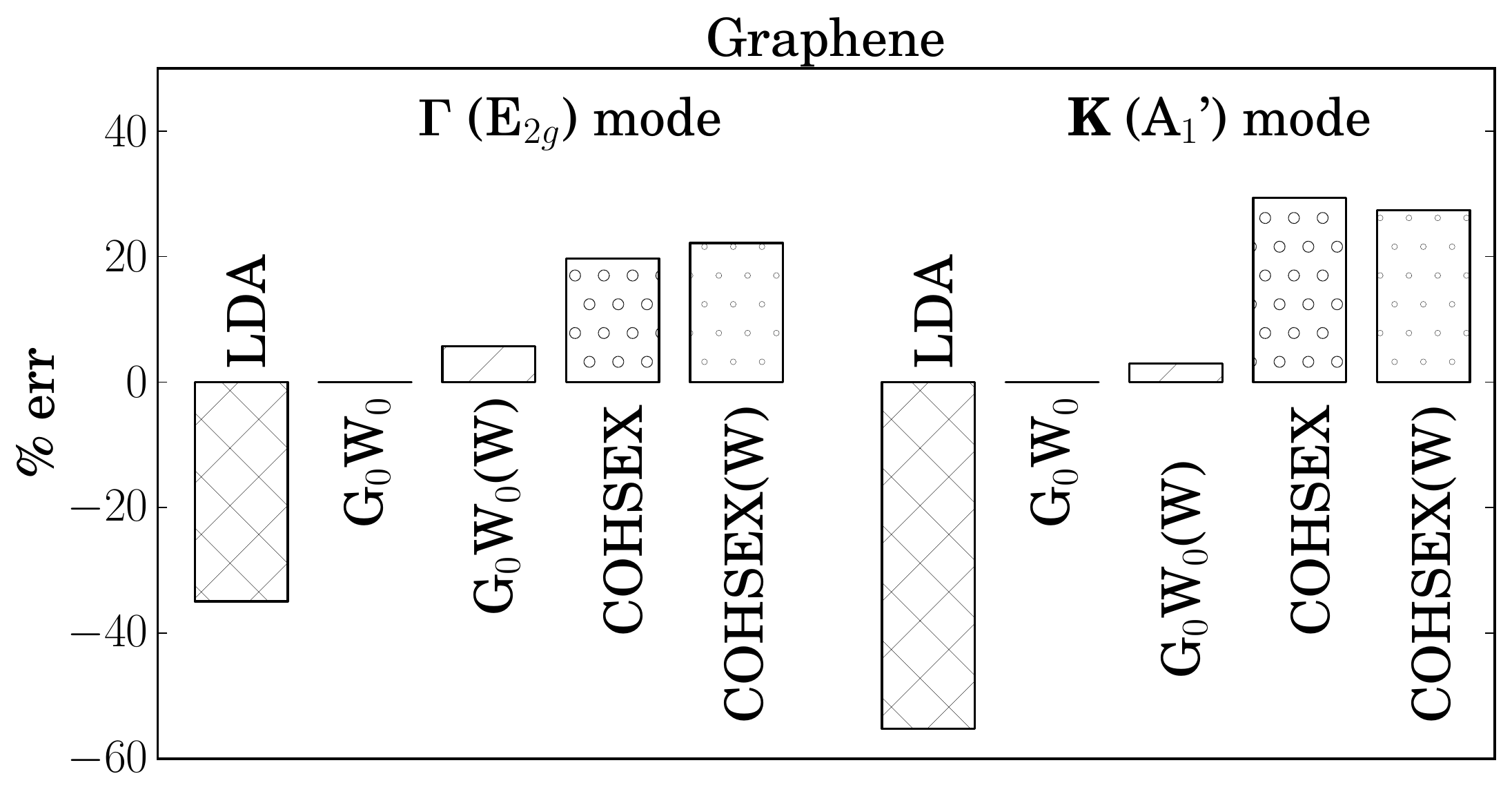}
  \caption{Relative error of the different approximations with respect to the $G_0W_0$ results for the electron-phonon coupling in graphene. }
  \label{figgraphene}
\end{figure}

\begin{figure}
 \includegraphics[width=0.8\linewidth]{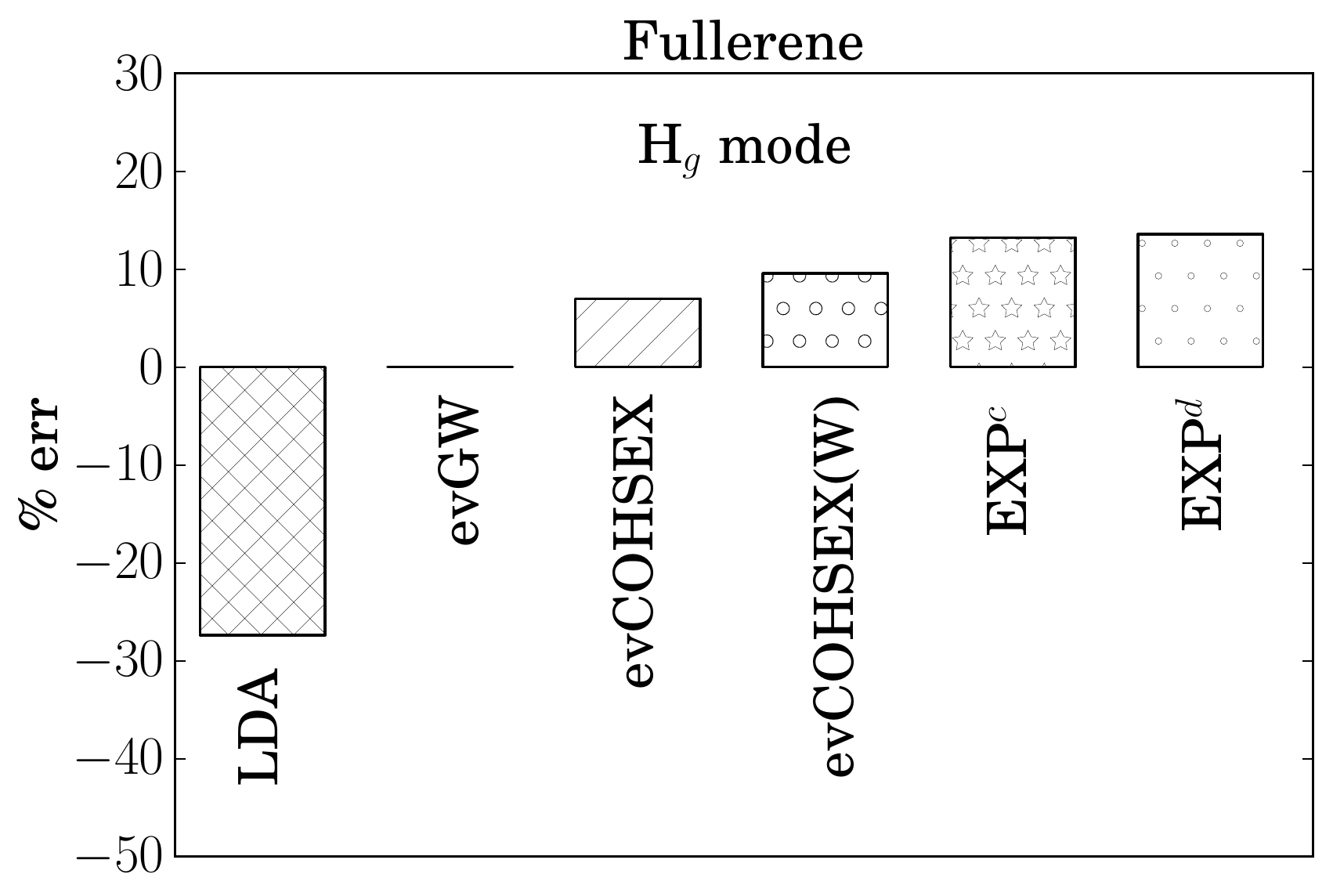}
 \caption{Relative error of the different approximations with respect to the ev$GW$ results for the $H_g$-related  $V^{ep}$ of
  equation (\ref{eq:Vep}). For convenience, a comparison to the experimental data is provided.}
   \label{figfullerene}
\end{figure}


We now explore approximations to the full $GW$ calculations performed in Refs.~\onlinecite{Faber11,Lazzeri08,Antonius14} and
reproduced here above. We start by the case of diamond. As compared  to the $G_0W_0$ calculations,  the static COHSEX approximation 
is shown to induce errors larger than 100$\%$ for the relevant electron-phonon coupling  energy to the  $\Gamma_{15c}$ and $X_{1c}$ states. In both cases, 
the static COHSEX approach strongly overestimates the electron-phonon coupling. A representation of the errors as compared to the 
$G_0W_0$ calculations is provided in Fig.~\ref{figdiamond}.

The case of graphene is of specific interest since, in great contrast to diamond and $C_{60}$, the fundamental gap reduces to zero in the equilibrium
geometry and opens upon phonon distortions. For the optical phonon mode at the zone-center, the COHSEX approximation leads again to an
increase as compared to the $G_0W_0$ value, even though smaller than in the diamond case with a 20$\%$ error. Such an error becomes 
larger for the ${\bf K}$-A$_{1}^{'}$ mode with a 29$\%$ increase of the EPC-related gap-opening rate. In the present graphene case,
the error induced by the COHSEX approach remains smaller than the  34$\%$ and 56$\%$ reduction observed upon using DFT-LDA 
instead of  $G_0W_0$. However, the close to 30$\%$ error observed for the COHSEX calculation of  $\langle D^2_{\bf K} \rangle$ is clearly 
significant (see Fig.~\ref{figgraphene}), questioning again the applicability of the static COHSEX formalism for EPC matrix elements calculations.

We now consider the $C_{60}$ case, where we compare ev$GW$ and evCOHSEX calculations. A global increase of about 7$\%$ from ev$GW$ 
to its static approximation is observed.  Besides the $H_g$(4) and $A_g$(1) modes showing very small couplings, the static COHSEX approximation
is observed again to systematically increase the coupling strength as compared to ev$GW$. \cite{cohsexc60}  This result is consistent with
what was observed for diamond and graphene, even though clearly the error appears to vary from one system to another.

\subsection{The constant screening approximation}

Following the above-mentioned analogy with the Bethe-Salpeter formalism, we finally test another approximation, that is
the constant screening approach. Namely,  we explicitly calculate the screened Coulomb potential for the undistorted 
structure and upon changing the positions of the ions along the (${\bf R}_{{\bf q}\nu}$) vibrational eigenmodes, 
we keep $W$ frozen to the undistorted value. Formally, this amounts to assuming that:
$\;  \partial  GW /  \partial \mathbf{u}^{{\bf q}\nu}  \approx (\partial  G /  \partial \mathbf{u}^{{\bf q}\nu} )W$.
Such an approach will be labeled $GW(W)$ or $COHSEX(W)$ in the following.

In the case of the fullerene, where calculations are performed within a Gaussian basis,  special care must be taken in the implementation 
 of such a constant-screening approximation. In the auxiliary basis, or resolution of the identity approach, the bare and screened 
Coulomb potentials are expressed in terms of an atom-centered auxiliary basis with the following relations:

\begin{eqnarray}
[W]_{\beta,\beta'} = \int \int d{\bf r} d{\bf r}' {\beta}({\bf r}) W({\bf r},{\bf r}') {\beta}'({\bf r}'), \\
W({\bf r},{\bf r}') = \sum_{\beta,\beta'}
   {\beta}({\bf r}) \left(  S^{-1} [W] S^{-1} \right)_{\beta,\beta'}  {\beta}'({\bf r}'),
\end{eqnarray}

\noindent where $S$ is the overlap matrix in the auxiliary basis. Using now the notation: $\underline W$ and
$\underline \beta$ for the screened Coulomb potential and the auxiliary basis for the slightly distorted
system, the assumption: $\; {\underline W}({\bf r},{\bf r}') \simeq W({\bf r},{\bf r}')$
leads straightforwardly to the condition:

\begin{eqnarray}
 [{\underline W}]_{{\underline \beta},{\underline \beta}'} \simeq S_{{\underline \beta}\beta} 
      [W]_{\beta,\beta'} S_{\beta' {\underline \beta}'},
\end{eqnarray}

\noindent where $S_{{\underline \beta}\beta}= < {\underline \beta}|\beta>$ is an overlap matrix between the
auxiliary bases for the perturbed and unperturbed systems, respectively.

For $C_{60}$ again, we  test the constant screening approximation at the COHSEX level only, namely comparing
evCOHSEX and evCOHSEX(W) calculations of the $V^{ep}$ energies.  While the plasmon-pole approach used 
for diamond and graphene is based on a fixed finite frequency pole value independent of the geometry, the real 
axis poles contribution to the correlation energy in the contour deformation approach  
(see Refs.~\onlinecite{Fiesta1,Farid})
implemented in the {\sc{Fiesta}} package changes from one structure to another, leading to difficulties when trying
to implement the constant-W approach within the frozen-phonon scheme.

The results of the constant-screening approximation are compiled in Tables~\ref{table1} and \ref{table2}. For $C_{60}$,
an excellent agreement is obtained within $evCOHSEX(W)$ compared to the corresponding evCOHSEX calculations. 
Comparing the total coupling, $evCOHSEX(W)$  and evCOHSEX agree within 1.5 \%.  

We now explore the constant-screening approximation for periodic diamond. Results are again compiled
in Table~\ref{table1}. For diamond, we observe a 3.4$\%$ and 7.5$\%$ difference between $G_0W_0(W)$ and $G_0W_0$ for the
$X_{1c}$ and $\Gamma_{15c}$ states, respectively. At the COHSEX level, the constant screening approximation yields errors of
the order of 7.0$\%$ and 10.2$\%$, respectively. In summary, while the COHSEX approximation was failing for diamond,
the constant-$W$ approximation leads again to very reasonnable results, with errors well within 10$\%$ at the $GW$ level.
As illustrated in Fig.~\ref{figdiamond}, this error is much smaller than the error induced by the LDA approximation.

We finally address the case of graphene.
Concerning the coupling with the zone-center optical mode, the constant-$W$ approach leads to a 5.7$\%$ and 2.1$\%$ 
error respectively when comparing $G_0W_0(W)$ to $G_0W_0$ and   COHSEX(W) to COHSEX. 
In the case of coupling with the zone-corner ${\bf K}$-A$_{1}^{'}$   phonon mode, 
the discrepancy is of 3$\%$ and 1.5$\%$ applying the constant-screening approximation to $GW$ and COHSEX 
approaches, respectively. Notice that the results for the constant-screening of the ${\bf K}$-A$_{1}^{'}$ phonon 
are affected by a numerical error of about 4$\%$, therefore the discrepancy could be slightly larger. 
Anyway this difference  is much smaller than the error induced by the standard DFT-LDA calculations, but also
much smaller than the error induced by the static COHSEX formalism as compared to the reference $GW$ calculation.

\section{Discussion}

Clearly, among the two approximations explored here above, the constant-screening stands as a much better approach
than the static COHSEX for the calculation of electron-phonon coupling matrix elements within the $GW$ formalism. 
Overall, the largest error induced by the static COHSEX approximation is larger than 100$\%$ 
in the diamond case, while it is reduced to 7$\%$ in the case of the constant-screening approximation to full
$GW$ calculations.

The static COHSEX approximation is known to always overestimate band gaps. As proposed in Ref.~\onlinecite{Janssen10}, a too large band gap 
should lead to an underscreening of the electron-phonon interaction and consequently to enhanced EPC matrix elements. This is consistent with the 
fact that, within Hartree-Fock or with increasing exact exchange in hybrid functionals, the EPC matrix elements are found to steeply increase. 
The Hartree-Fock $\langle D^2_{{\bf \Gamma}} \rangle$ and $\langle D^2_{{\bf K}} \rangle$  coupling constants in graphene were found to be 
about 5 and 30 times  larger, respectively, as compared to their $GW$ analog, \cite{Lazzeri08} while in $C_{60}$ increasing the percentage 
of exact exchange from 20$\%$ to 30$\%$ in hybrid functionals was found to enhance the $V^{ep}$ energy by about 15$\%$.  \cite{Janssen10}
Such an interpretation matches the observation that in graphene and $C_{60}$, the COHSEX EPC coupling constants are larger than their corresponding
$GW$ reference. 
In the case of diamond, where we are concerned with the second-order derivatives, the effect of assuming a static approximation may be more 
difficult to interpret.

An important observation further is that the Coulomb-hole term is a local potential in the static COHSEX approximation, washing out spatial 
local-field effects, a observation that may potentially allow to understand the failure of the COHSEX approximation to reproduce accurately
the evolution of the band structure with phonon deformation. In any case, the large variations of the induced errors from diamond to graphene
and $C_{60}$ is still to be understood, and we will just stand here on the observation that the static COHSEX approximation cannot be trusted to
improve on DFT-LDA calculations. 

Concerning the neglect of the gradient of $W$ with respect to the ionic positions, such an approximation was already tested by Ismail-Beigi 
and Louie in the context of excited state ionic forces within the Bethe-Salpeter formalism, \cite{Beigi03} showing good agreement with explicit 
finite-difference "exact" BSE calculations for small CO and NH$_3$ molecules. 
It is commonly assumed that the $GW$/BSE approach is much more resistant to approximations on $W$ as compared 
to the $GW$ approach for (charged) excitations. This is due to cancellations of errors between the electron-electron and electron-hole interactions.
Namely, any error introduced in $W$ is expected to affect excitonic interactions and quasiparticle gaps  in opposite ways. Clearly,
the present $GW$ study of the variations of a given quasiparticle energy with respect to ionic positions cannot benefit from such
cancellation of errors. Still, the constant-screening approximation turns to be a reliable approach to save on computational cost.

An important consequence of the present findings is that once the screened Coulomb potential $W({\bf r},{\bf r}'; \omega)$ is built for 
the equilibrium geometry, the calculation of the variations  of the quasiparticle energies with respect to the perturbation ($\lambda$) only 
requires the evaluation of  the variations of the Green's function $G$ with respect to the perturbation. This can be  performed
within standard DFPT techniques, at least in the case of non-self-consistent $G_0W_0$ calculations where the Green's function assumes
an explicit form in function of the input DFT eigenstates. This may invite, for systems such as $C_{60}$, to use $G_0W_0$ calculations 
starting from DFT eigenstates obtained with hybrid functionals, which have been shown to be a better starting point for 
finite-size systems. \cite{Marom,Bruneval13,Korbel14,Boulanger14} 

\section{Conclusions}

We have explored two approximations for calculating self-energy gradients of interest for electron-phonon coupling, namely the two  simplifications 
commonly used in the $GW$/Bethe-Salpeter calculations, that is the static COHSEX and the constant screening approximations. We explored these
approaches in the case of diamond, graphene and $C_{60}$. Our findings suggest that the COHSEX approximation cannot be trusted to improve on
the DFT-LDA values as clearly illustrated in particular in the case of  diamond. On the contrary, the constant screening hypothesis, namely assuming that
$W$ remains to first order constant with respect to small ionic displacements, seems to be reliable, with a  discrepancy no larger than 10$\%$, even in
the difficult case of graphene where the phonon perturbation dramatically affects the Dirac cone and the (semi)metallic nature of the 
graphene sheet. Even though a deeper understanding of the origin of the specific difficulties uncountered by the static COHSEX approximation
would allow to better rationalize the validity  and limits of the tested approximations,  
the present results offer promising perspectives to carry on such many-body evaluations of the electron-phonon 
coupling gradients with much reduced computer cost on realistic systems, including the study of periodic systems with arbitrary wave vector perturbation.  
Further studies are however required on a larger  set of systems and physical observables in order to better assess the interest, with respect to common DFT 
calculations, of using the  $GW$ approach,  and its various  approximations,  for calculating self-energy gradients with respect to ionic positions. 

\textbf{Acknowledgments.} 
P.B. warmly acknowledges G. Antonius for providing the diamond phonon eigenmodes used in this study and for guiding our comparison with
Ref.~\onlinecite{Antonius14}. C. A. acknowledges Daniele Varsano for the correction of some bugs in the Yambo code.  
C.F. is indebted to the French CNRS and CEA for Ph.D funding.
P.B. acknowledges a postdoctoral fellowship from the French  National Research Agency under Contract 
No. ANR-2012-BS04 PANELS. Computing time has been provided by the ``Curie"  national GENCI-IDRIS supercomputing 
center under contract No. i2012096655 and a PRACE European project under contract No. 2012071258.
X.B. acknowledges illuminating discussions with Steven G. Louie and G. Antonius.


\begin{thebibliography}{150bliography}
\bibitem{noteph}
One has to note however  that  the impact of the functional  on the vibrational  eigenmodes used to calculate the EPC matrix elements 
was found to be marginal, the main effect coming from the framework  within which the quasiparticle energies are calculated.
\bibitem{notegraphene}
    Notice that in Ref.~\onlinecite{Lazzeri08} the interlayer distance between the graphene sheets is 15 a.u. and not 20 a.u. as it is wrongly reported in the paper.
\bibitem{cohsexc60}
We provide for information in Table~\ref{table2} the results associated with the (single-shot) COHSEX approach, 
starting from DFT-LDA eigenstates. Clearly, the EPC coupling energies approach from below the ev$GW$ value with 
an error not larger than that induced by the evCOHSEX technique. This stems from a cancellation of errors between
the COHSEX approximation and the lack of self-consistency which leads to overscreening when building $W$ directly 
from the DFT-LDA spectrum.
\end{thebibliography}
\end{document}